\documentclass[twocolumn, deluxetables]{aastex631}
\usepackage{amsmath}
\usepackage{multirow}
\usepackage{graphicx}
\usepackage{xspace}

\newcommand{\rxte}{{\it RXTE}\xspace}

\newcommand{\ginga}{{\it GINGA}\xspace}
\newcommand{\maxi}{{\it MAXI}\xspace}

\newcommand{\swift}{{\it Swift}\xspace}
\newcommand{\nicer}{{\it NICER}\xspace}

\newcommand{\nustar}{\textit{NuSTAR}\xspace}

\newcommand{\gs}{GS 1826-238\xspace}

\graphicspath{{./}{figures/}}

\shorttitle{GS 1826-238 in stray light}
\shortauthors{Yun et al.}

\begin{document}

\title{Revealing the spectral state transition of the Clocked Burster, \\  GS 1826-238 with NuSTAR StrayCats}

\author{S.~B.~Yun}
\affil{Department of Astronomy, The University of Michigan, 1085 S. University Ave., Ann Arbor, MI, 48109, USA}

\author[0000-0002-1984-2932]{B.~W.~Grefenstette}
\affiliation{Cahill Center for Astronomy and Astrophysics, California Institute of Technology, Pasadena, CA 91125, USA}

\author[0000-0002-8961-939X]{R.~M.~Ludlam}
\affiliation{Department of Physics \& Astronomy, Wayne State University, 666 West Hancock Street, Detroit, MI 48201, USA}

\author[0000-0002-4024-6967]{M.~C.~Brumback}
\affil{Department of Astronomy, The University of Michigan, 1085 S. University Ave., Ann Arbor, MI, 48109, USA}

\author[0000-0002-5341-6929]{D.~J.~K.~Buisson}
\affil{Independent}

\author[0000-0003-4216-7936]{G~Mastroserio}
\affiliation{INAF-Osservatorio Astronomico di Cagliari, via della Scienza 5, I-09047 Selargius (CA), Italy}

\author[0000-0002-8403-0041]{S.~N.~Pike}
\affiliation{Center for Astrophysics and Space Sciences, University of California, San Diego, CA 92093, USA}




\begin{abstract}
We present the long term analysis of \gs, a neutron star X-ray binary known as the ``Clocked Burster", using data from \nustar StrayCats. StrayCats, a catalogue of \nustar stray light data, contains data from bright, off-axis X-ray sources that have not been focused by the \nustar optics. We obtained stray light observations of the source from 2014-2021, reduced and analyzed the data using nustar-gen-utils Python tools, demonstrating the transition of source from the ``island" atoll state to a ``banana" branch. We also present the lightcurve analysis of Type I X-Ray bursts from the Clocked Burster and show that the bursts from the banana/soft state are systematically shorter in durations than those from the island/hard state and have a higher burst fluence. From our analysis, we note an increase in mass accretion rate of the source, and a decrease in burst frequency with the transition.

\end{abstract}



\section{Introduction} \label{sec:intro}


Low-mass X-ray binaries (LMXBs) are X-ray sources which consist of a compact object such as a black hole (BH) or a neutron star (NS) which accretes material from a main sequence companion star, primarily via Roche-lobe overflow from the companion star. If the system contains a neutron star (NS), then the X-ray emission typically arises from thermal emission from the NS surface and in the accretion disk. This thermal emission passes through a region of nonthermal electrons known as the corona and is reprocessed via Comptonization into a power-law emission tail. The discovery of many of these sources by the \ginga\ satellite in the 1980s led to a multitude of discoveries, including the production of Type I X-ray bursts (thermonuclear bursts occurring when accreted material on the neutron star surface reaches a critical density) and a zoo of LMXB behaviors \cite[see, e.g.][]{Lewin_1993}. In the last few years, studies of these sources with \nustar\ and \nicer\ have shown that understanding the physics of LMXBs can lead to fundamental measurements of the equation of state of NSs \citep[e.g.][]{Ludlam_2022}.

Discovered in 1988 by \ginga\ \citep{Makino_1988}, \gs\ was originally considered a black hole candidate. Measurements of the high energy curvature in the spectrum \citep{Strickman_1996} argued for a NS nature of the compact object, which was later confirmed through detection of numerous Type I X-ray bursts \citep{Ubertini_1999}. The bursts recurred on such regular roughly 6-hour intervals that \gs\ earned the monicker ``The Clocked Burster" and became a prototypical source for observations of recurring thermonuclear bursts. Over the next 25 years, the source remained in the low/hard state with the spectrum characterized by various flavors of thermal Comptonization (e.g., \texttt{cutoffpl, comptt, comptb}, etc) with a high energy cutoff $>$ 50 keV \citep[e.g.][]{Strickman_1996, Barret_2000, Cocchi_2010}. Observations using \rxte\ enabled detailed analyses of the relation between the mass accretion rate and the burst recurrence  time \citep{Galloway_2004}. This has enabled fundamental tests of nuclear physics and the nuclear equation of state (EOS) in the NS.


There has been a claim of excess emission near the Fe line complex \citep{Barret_2000, Ono_2016}, which was associated with neutral Fe emission arising from reflection off of cold material near the source. However, this excess emission has not generally been reported in other hard state observations of the source, though (when reported) the equivalent width of the $\sim$0.5 keV broad Fe line was relatively weak at 50 eV.

Starting in 2014, the source began to transition away from the hard state after more than 25 years. This triggered a focused observation from \nustar, where the source showed some evidence for a changing accretion geometry and a photospheric radius expansion (PRE) Type-I X-ray burst providing a distance of roughly 5.7 kpc \citep{Chenevez_2016} (We adopt this as the distance to the source for the remainder of this paper).

Shortly after this ``soft episode", the source returned to its hard state. Observations using \swift-XRT during this period indicate that the source remained in an ``intermediate atoll" state \citep{Ji_2017} and had not yet fully transitioned to the soft state. Presumably, the change in the observed behavior is correlated to a change in the mass accretion rate onto the NS. A crucial missing data point here is the behavior of the Type I X-ray bursts during the state transition.

Other than short snapshots from \nicer, one focused observation with \nustar, and a pair of observations using \textit{ASTROSAT} \citep{agrawal_2022}, until 2022 the source has not been intentionally observed since it started its transition. Fortunately, \gs is nearby to several sources with recurring outbursts (e.g., V4641 Sgr). This has resulted in a large set of additional serendipitous ``stray light" observations found in the \nustar StrayCats database \citep{Grefenstette_2021, Ludlam22a}. These are typically longer duration than the focused observations, and allow us to track the source behavior throughout the state transition.

This paper is structured as follows: in \S \ref{sec:longterm} we discuss the long-term behavior of \gs as measured by high-energy all-sky monitors and put the various pointed observations and the StrayCat observations in context; in \S \ref{sec:results} we present an analysis of the source both before and after the state transitions and investigate the behavior of the Type I X-ray bursts, and in \S \ref{sec:discussion} we discuss the slow transition and change in bursting behavior.

\vfill 

\section{\gs long-term behavior}
\label{sec:longterm}

 We downloaded the publicly available \maxi \footnote{http://maxi.riken.jp/top/index.html} and \swift \footnote{https://swift.gsfc.nasa.gov/results/transients/} data from their respective public websites. We have not done any additional reprocessing of the data and thank both the \maxi and \swift teams for their dedication to open access science. The \maxi data span MJD 55050 through MJD 59661 at the time of retrieval. We note that as of this writing the publicly available \swift data are limited to data only taken after MJD 59289 due to internal processing. Fortunately, we had previously retrieved the \swift data for \gs so our data set extends back to MJD 53415. We annotated the long-term lightcurves showing the timing of the published \nustar observation \citep{Chenevez_2016}, the \nustar stray light observations from the StrayCats catalog \citep{Grefenstette_2021}, and the \textit{ASTROSAT} targeted observations.

\begin{figure*}[ht]
\centering
\includegraphics[width=0.75\textwidth]{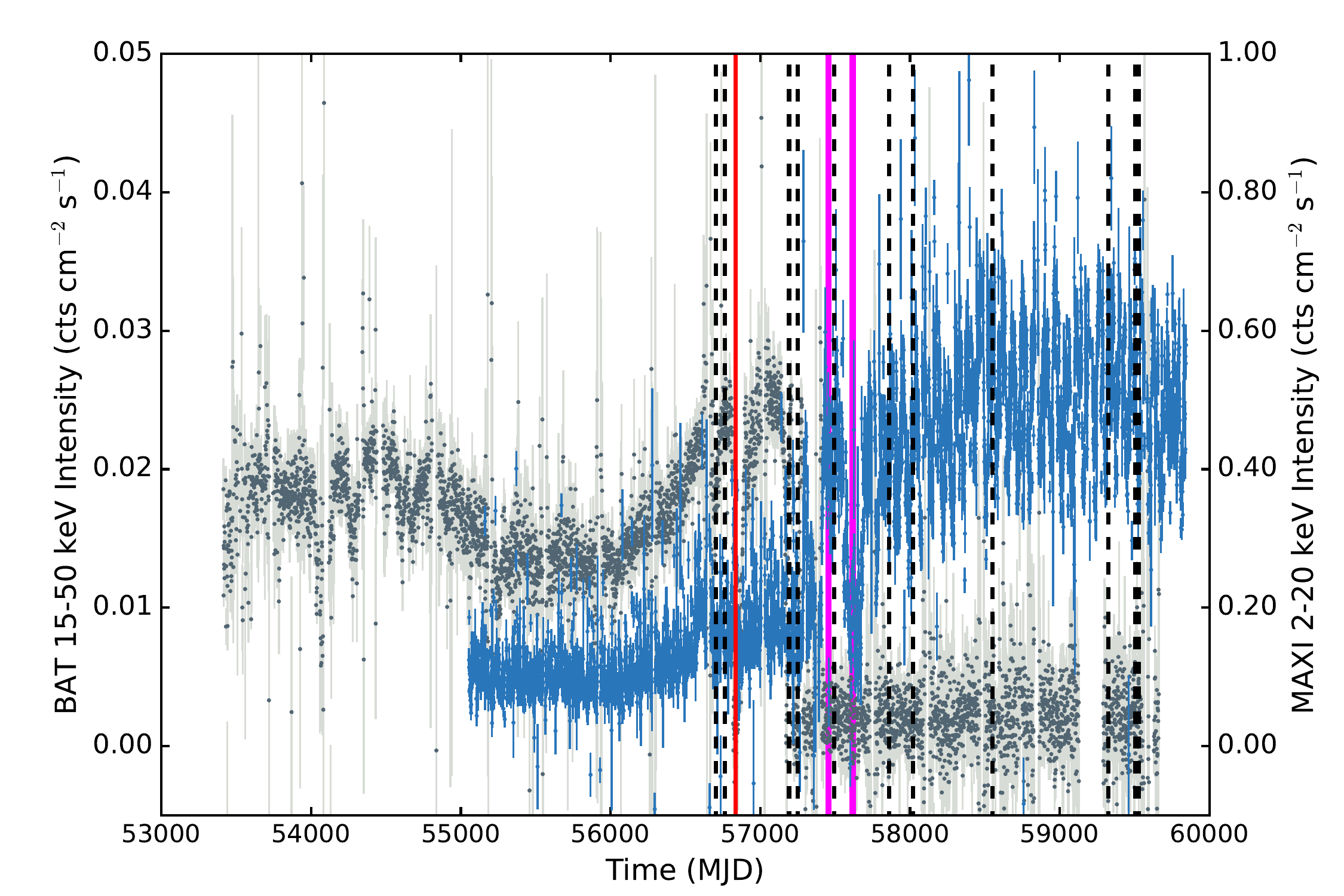}
\vspace {1mm}
\caption{The long-term \maxi (blue) and \swift-BAT (grey) lighcurve for \gs. Vertical lines correspond to: Red, \nustar focused observation; Magenta, ASTROSAT observations; dashed black, \nustar stray light observations listed in \ref{table:Observation}}
\label{fig:longterm}
\end{figure*}

\begin{figure}[ht]
\centering
\includegraphics[width=0.45\textwidth]{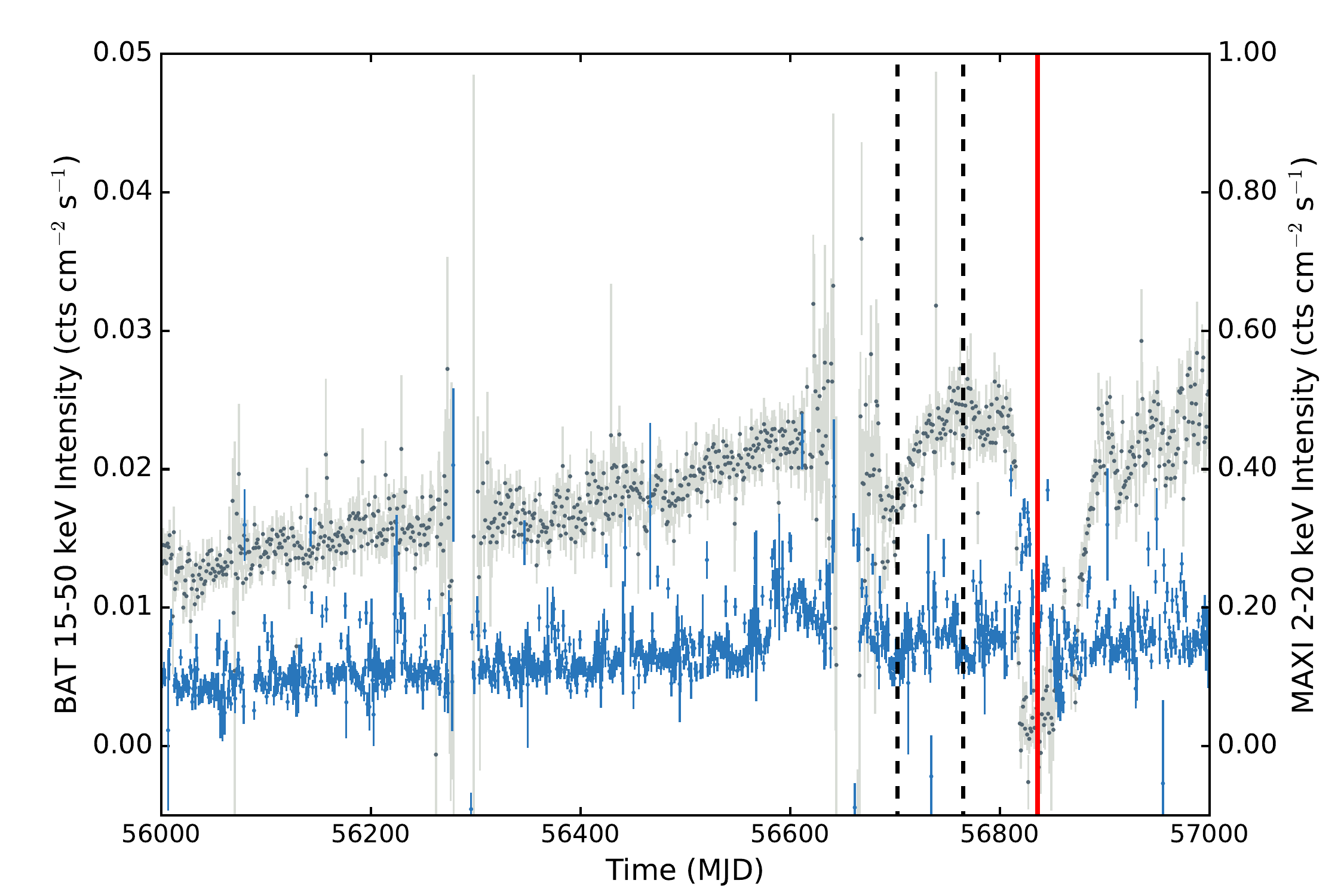}
\includegraphics[width=0.45\textwidth]{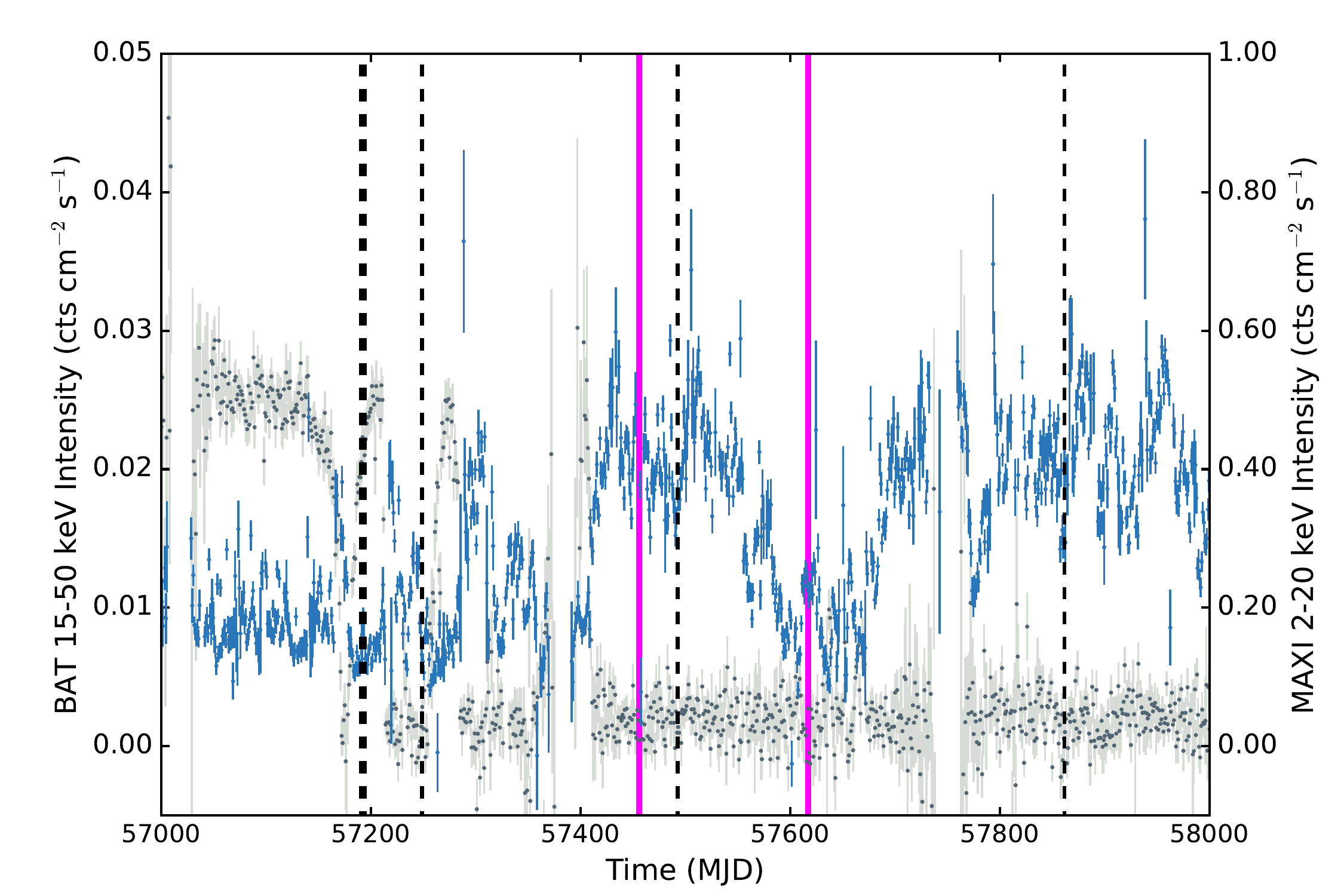}
\vspace {1mm}
\caption{Same as Fig 1, but showing zoomed in views on the long-term light curve showing the initiation coronal variations (top) and the transitional period (bottom).}
\label{fig:hard}
\end{figure}

The long-term lightcurve (Fig \ref{fig:longterm}) shows the evolution of the source from the hard state to the soft state. By MJD 55000 ($\sim$mid 2009) the source was starting to show signs of evolution in its behavior in the \swift lightcurve, showing a rise in both the hard and soft flux heading into the 2010s. In 2014 the source showed the first evidence for a collapse of the hard band flux (Fig \ref{fig:hard}, top panel). 

Variations in the hard flux lasted until roughly MJD 57100 (March 2015), when the hard flux completely disappeared, while the soft flux showed significant variability over a multi-year period (Fig \ref{fig:hard}, bottom panel). The \textit{ASTROSAT} observations show evidence for a classical banana ``soft state" for the source similar to the 2014 soft episode, with the emission dominated by an optically thick corona with a low electron temperature \citep{agrawal_2022}.

\begin{figure}[ht]
\centering
\includegraphics[width=0.45\textwidth]{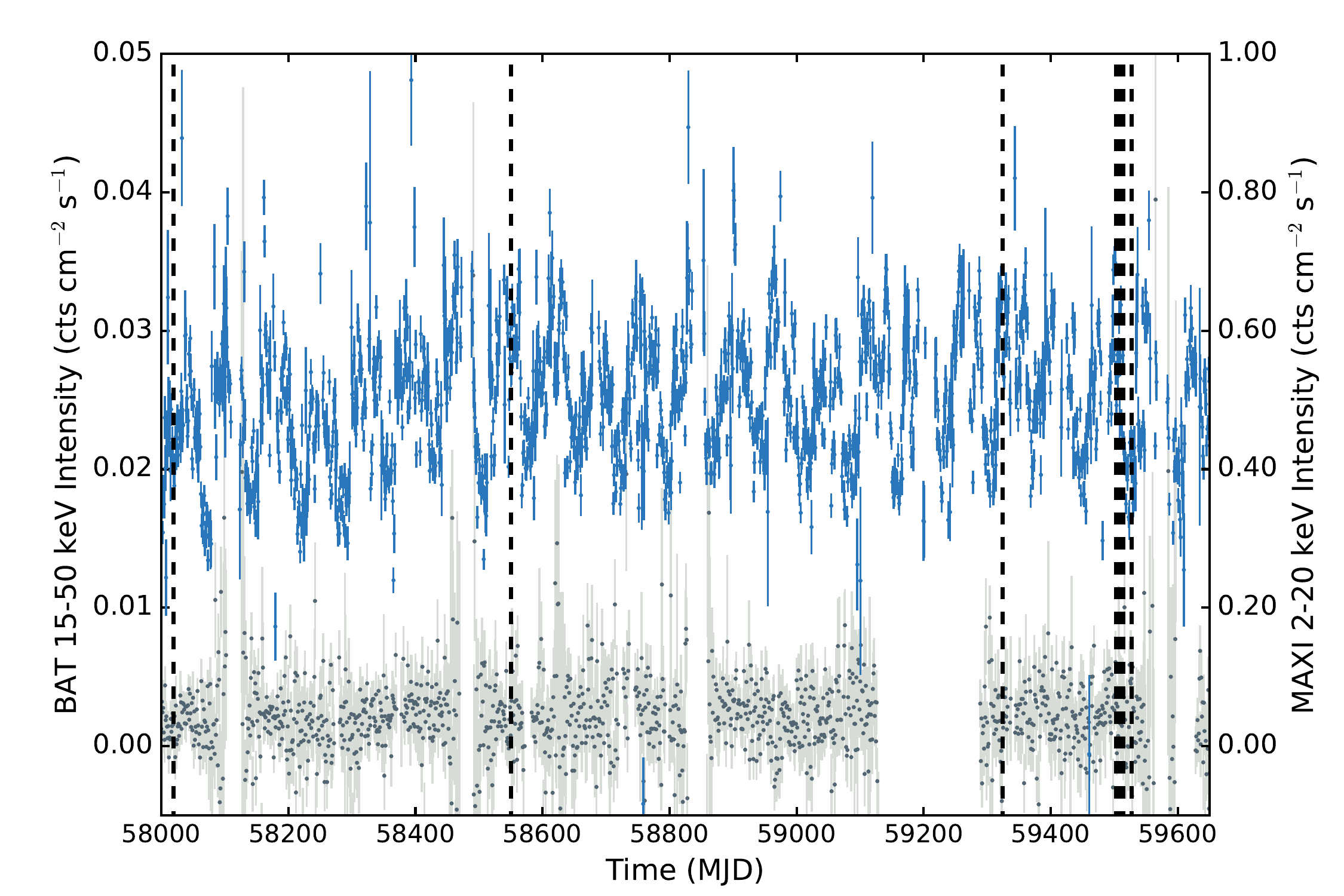}
\vspace {1mm}
\caption{Same as Fig 1, but showing the ``soft state" period. We note here that the periodicity seen in the \maxi data at a period of roughly 70-days is artificial and corresponds to the orbital precessional period of the International Space Station.}
\label{fig:soft}
\end{figure}

From MJD 57900 (mid-2017) to the present, the source has further evolved to a stable, soft state (Fig \ref{fig:soft}). This corresponds to the period when NICER observations indicate the presence of mHz quasi-periodic oscillations (QPOs) and indicates change in the accretion rate producing unstable burning on the surface of the NS, causing the QPO \citep{Strohmayer_2018}.

\section{StrayCats Observations and Data Reduction}
\label{sec:obs}

There have been 15 \nustar StrayCat observations of \gs between February 2014 and November 2021 (Table \ref{table:Observation}). We separate these into two clear categories of ``hard" observations (Observations 1-4) and ``soft" observations (Observations 5-15) that occur after the coronal collapse of the system (as measured by the drop in the \swift-BAT rate).

\begin{figure}[ht]
\centering
\includegraphics[scale=0.3]{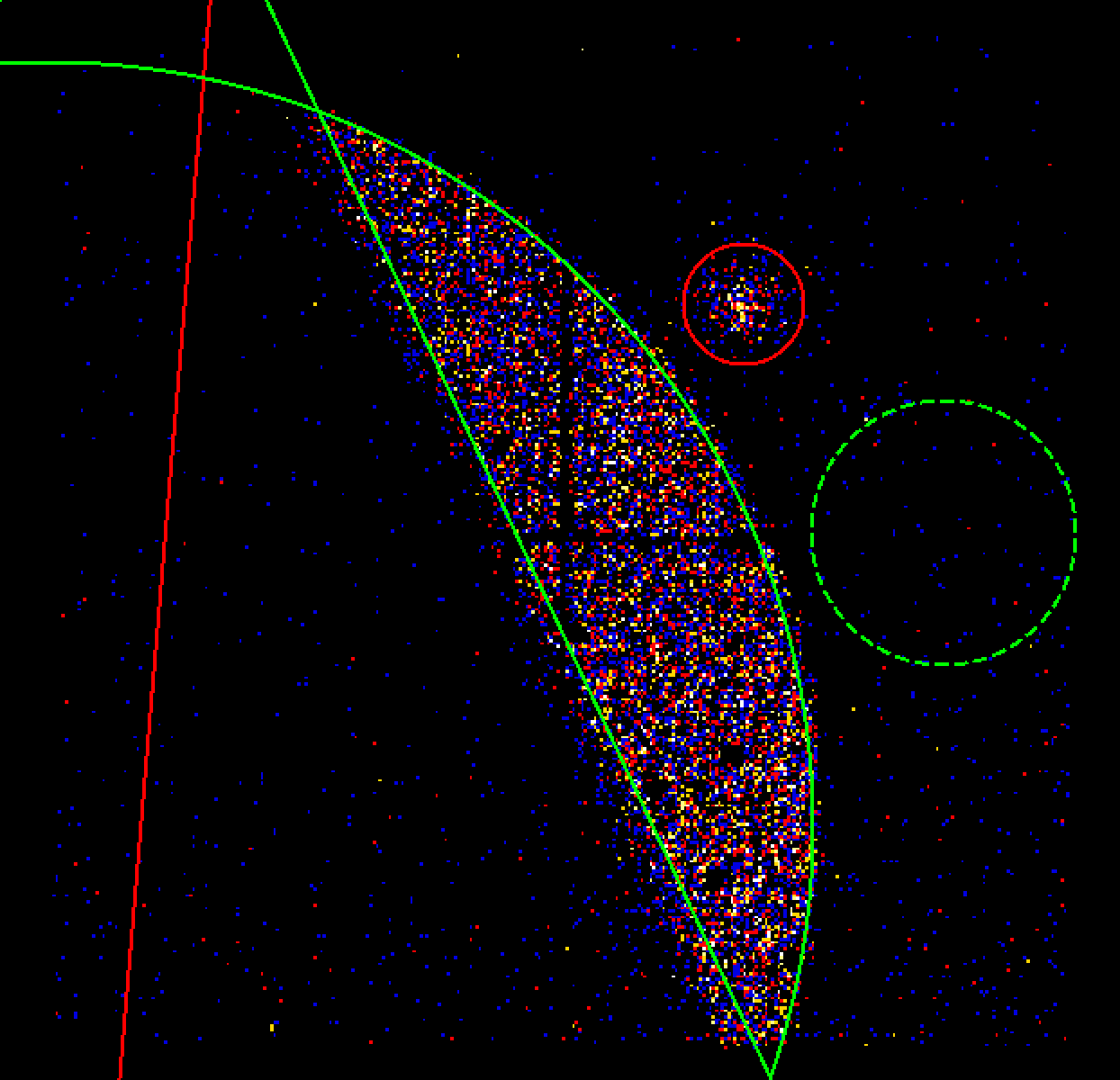}
\vspace {1mm}
\caption{A stray light observation of \gs \ (SeqID: 30101053002). The stray light region is defined by the green solid lines. The background region (green dashed lines) does not overlap with the straylight region or the focused target source (red). The source region is excluded from the stray light region if there is an overlap.}
\label{fig:straylight}
\end{figure}

\begin{table*}[th]
\caption{\gs \ \texttt{StrayCats} Observations}
\begin{footnotesize}
\begin{center}
\begin{tabular}{cccccccc}
Obs \# & Sequence ID &   Time  &   MJD   &   FPM   &   Exposure (ks)   &   Area ($cm^2$)  & \# Type 1 Bursts \\
\tableline
\tableline
1    & 80002012002 & 2014-02-14T00:36:07 &   56702.0   &   A   &   24.05   &   1.84   &   2   \\
2    & 80002012004 & 2014-04-17T22:46:07 &   56765.0   &   A   &   26.42   &   2.30   &   3   \\
3    & 30101053002 & 2015-06-17T16:06:07 &   57190.7   &   A   &   131.3   &   2.71   &   14  \\
4    & 30101053004 & 2015-06-21T07:11:07 &   57194.3   &   A   &   51.52   &   2.56   &   3   \\
\tableline
5A   & 90102011002 & 2015-08-14T12:21:08 &   57248.5   &   A   &   30.65   &   1.77   &   3   \\
5B   &      -      &          -          &      -      &   B   &   30.60   &   3.39   &   2   \\
6    & 60160692002 & 2016-04-14T18:26:08 &   57492.8   &   B   &   21.78   &   1.66   &   0   \\
7    & 10202005002 & 2017-04-18T13:06:09 &   57861.6   &   A   &   156.5   &   2.38   &   4   \\
8    & 10202005004 & 2017-09-23T08:36:09 &   58019.4   &   B   &   155.3   &   8.71   &   2   \\
9    & 80460628002 & 2019-03-08T20:21:09 &   58550.9   &   B   &   41.05   &   1.65   &   0   \\
10A* & 90701314002 & 2021-04-20T11:16:09 &   59324.5   &   A   &   36.28   &   0.20   &   0   \\
10B* & 90701314002 &          -          &      -      &   B   &   36.28   &   0.13   &   0   \\
11A  & 80702324002 & 2021-10-15T11:01:09 &   59502.5   &   A   &   18.04   &   1.28   &   0   \\
11B  &      -      &          -          &      -      &   B   &   17.97   &   1.38   &   0   \\
12A  & 80702324004 & 2021-10-19T13:11:09 &   59506.6   &   A   &   19.16   &   1.66   &   0   \\
12B  &      -      &          -          &      -      &   B   &   19.06   &   1.48   &   1   \\
13A  & 80702324006 & 2021-10-22T08:46:09 &   59509.4   &   A   &   17.47   &   1.38   &   1   \\
13B  &      -      &          -          &      -      &   B   &   17.39   &   1.30   &   1   \\
14A  & 80702324008 & 2021-10-26T23:56:09 &   59514.0   &   A   &   19.95   &   1.74   &   0   \\
14B  &      -      &          -          &      -      &   B   &   19.82   &   1.45   &   0   \\
15A  & 80702324009 & 2021-11-09T12:51:09 &   59527.5   &   A   &   20.12   &   1.77   &   0   \\
15B  &      -      &          -          &      -      &   B   &   20.00   &   1.46   &   0   \\
\tableline
\tableline
\end{tabular}
\vspace*{\baselineskip}~\\
\end{center} 
\tablecomments{The set of Sequence IDs above the solid line are observations during the hard spectral state whereas those below the line are observations of the soft state. Observations marked with an asterisk (*) were not used in the analysis due to low Stray Light area.}
\end{footnotesize}
\label{table:Observation}
\end{table*}

We processed and analysed all of the data using HEASoft v6.29c, NuSTARADS v2.1.1, NuSTAR CALDB v20211221 and the \texttt{nustar-gen-utils}\footnote{\url{https://github.com/NuSTAR/nustar-gen-utils}} package. All the observations were first reprocessed via \texttt{nupipeline}.

Stray light sources are observed in various shapes and sizes along the field of view (FoV). The stray light source regions were created using DS9, based on the aperture stop shadow projected onto the focal plane. Background regions were defined from the adjacent regions, avoiding overlap with both the stray light region and the target of the focused observation. The image shown in Figure \ref{fig:straylight} shows an example of the stray light and the background region used for analysis. The ``Stray Light Wrapper" scripts from \texttt{nustar-gen-utils} were used to produce the relevant high-level products used for analysis. We screened the observations and excluded those with a stray light area of less than 1 $\rm cm^2$ for our analysis as they were deemed to not have enough source counts. 

\section{Analysis and Results} 
\label{sec:results}

\subsection{HR Diagram}

As a preliminary look at X-Ray timing and its spectral behavior, we plotted a Hardness Ratio (HR) - Intensity diagram of \gs\ using light curves binned at 500s per bin. We used a hardness ratio based on \cite{Hasinger_1989} of the ``hard" 6-20 keV band pass and the ``soft" 3-6 keV band pass for each data point on the light curves. The HR was plotted against the total 3-20 keV intensity per area given that the stray light regions covers different sized regions on the focal plane for each observation (Table \ref{table:Observation}).

\begin{figure}[ht]
\centering
\includegraphics[scale=0.5]{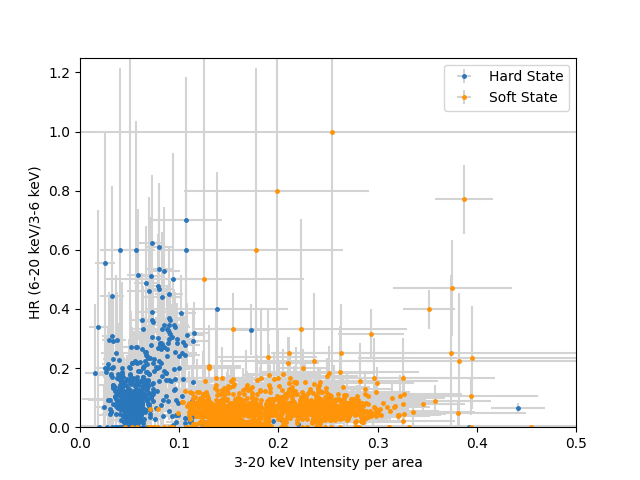}
\vspace {1mm}
\caption{The X-Ray Hardness Ratio vs Intensity diagram of \gs. The corresponding error bars for each data point are in grey. }
\label{fig:hr_diagram}
\end{figure}

Figure \ref{fig:hr_diagram} shows the resultant HR-Intensity diagram. It is possible to see two distinct states of atoll sources: the ``island" state reflecting the hard spectral state and the soft ``banana" state. We note that the data points constituting the island state are mostly from observations prior to 2016, while the soft banana state comprised primarily of data points from observations after 2016, which is analogous to the time period when \gs\ transitioned into a persistent soft state on the Maxi-BAT light curve. 

\subsection{Persistent Spectrum}
The X-Ray spectral fits were made using XSPEC v12.12.0. The fits were made across the 3-20 keV band, due to the source falling below background level at energies $>$20 keV. The quality of fits were measured using C-statistics and all error values quoted throughout the paper represent 1$\sigma$ uncertainties. 

We have chosen one representative spectrum from the hard and soft spectral states made based on the exposure time, count rates, and the amount of illuminated detector area. Observations with absorbed stray light \citep[see][]{madsen_observational_2017} or those with solar flares were excluded from the selection due to their complications in background modelling. We selected Obs 1 for the hard spectral state and Obs 8 for the soft spectral state. Since the source is bright, but soft, we model the the background using \texttt{nuskybgd}\footnote{\url{https://github.com/NuSTAR/nuskybgd-py}} to produce background models that we simultaneously fit to the data. We did not exclude time intervals when the source was undergoing Type 1 X-Ray bursts as their duration is short compared to the total exposure time.

For both states we first fit the data and then used the \texttt{emcee} implementation in \textsc{Xspec} to estimate the confidence intervals once we confirmed by eye that the solution had stabilized. In all cases we froze the neutral absorption to be $0.4 \times 10^{22} \: \rm{cm}^{2}$ \cite{intzand_1999} since \nustar is not sensitive to such low levels of absorption on its own.

\begin{figure*}[th]
\centering
\begin{minipage}{.48\textwidth}
    \centering
    \includegraphics[width=1.0\textwidth]{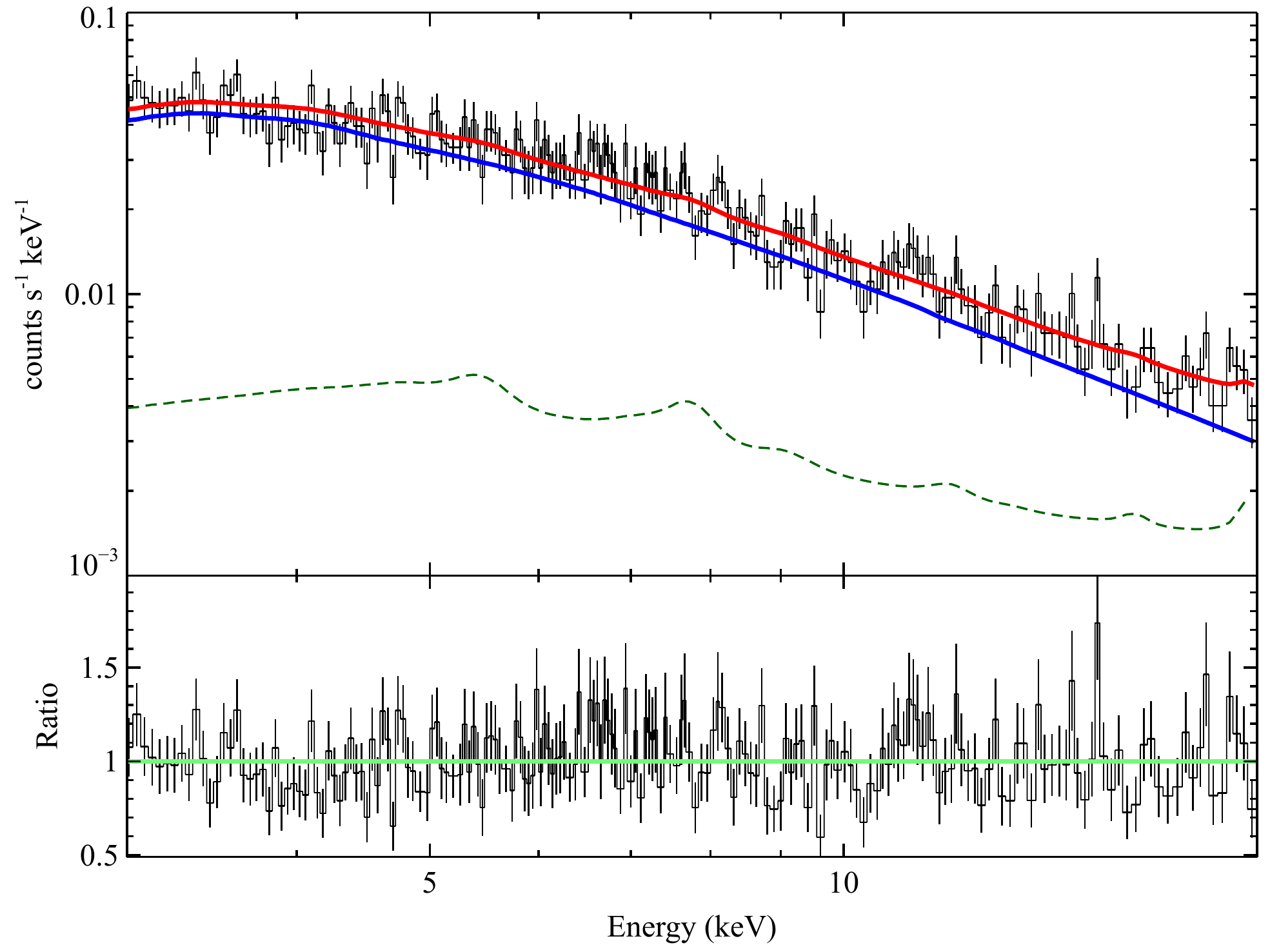}
\end{minipage}%
\hfill
\begin{minipage}{.48\textwidth}
    \centering
    \includegraphics[width=1.0\textwidth]{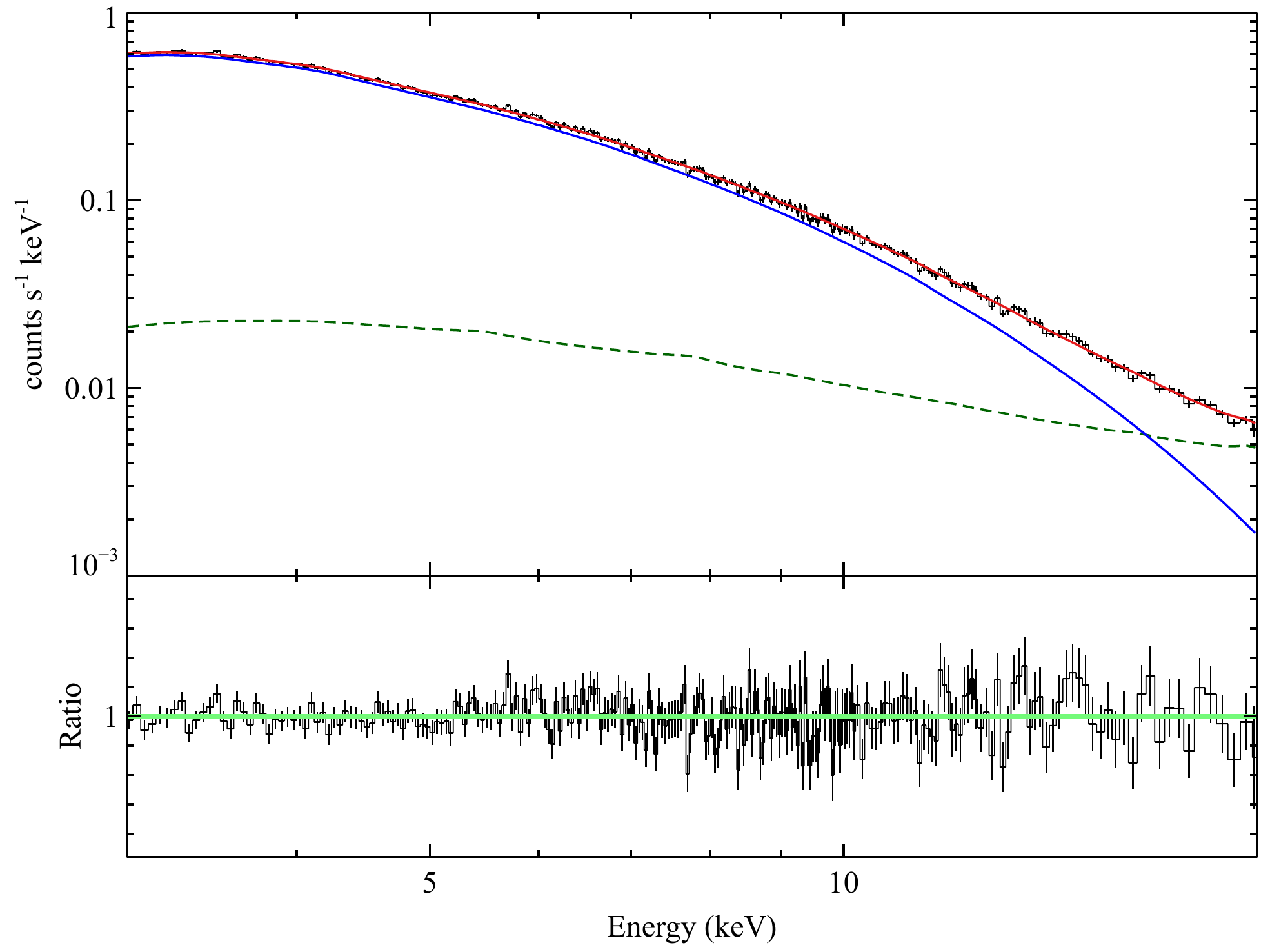}
\end{minipage}
\caption{The persistent spectrum of of \gs \ during the hard (\textit{Left}) and soft (\textit{Right}) epochs (see Section 4.2) fitted with a single Comptonization model. The \texttt{nuskybgd} model components are shown dashed green lines, while the source model is shown in blue solid lines. The bottom panels show the residuals to the fitted models for the spectra. The fits were made using unbinned spectra, and were binned to 5 and 20 counts for hard and soft state respectively for visual representation.} 
\label{fig:spectral_fits}
\end{figure*}

\subsection{Hard State Spectrum}

We fit the hard state spectrum with a single Comptonization model using \texttt{tbabs*(cflux*compTT)} in \textsc{Xspec}. While we were able to obtain a reasonable fit to the data allowing the seed and electron temperature, the optical depth, and the flux to vary, many of the parameters were highly correlated and poorly constrained. This is likely due to the fact that the photon seed temperature is below the \nustar band-pass of 3 keV (so we do not detect a significant low-energy roll-over) and the plasma temperature is above the point where the background starts to dominate the spectrum. Using a fixed seed temperature of 0.6 keV we find a best-fit plasma temperature of 17 keV. However, using the Xspec \texttt{error} command to explore the allowed parameter space we find that we can only place a lower limit of 8 keV on the plasma temperature. This is consistent with literature values for the spectral model that dominates in the \nustar band \citep[][e.g.,]{Thompson_2008, Chenevez_2016}. For simplicity and to compare with previous work we freeze the plasma temperature to 20 keV.

Using the fixed plasma temperature, we then allow the seed temperature to vary. While the best fit value is 0.6 keV, the \texttt{error} run indicates that we can only set the temperature of the seed photons to be less than 0.8 keV.

\subsection{Soft State Spectrum}

In the soft state, the spectrum of the source is qualitatively different. We can still obtain a reasonable fit with the same model, though the plasma temperature has dropped dramatically and the optical depth has increased. These are all indicative of a classical transition between the ``island" and ``banana" states in an atoll source. The results of the fit are given in Table \ref{table:spectrum}.


\begin{table}[ht]
\caption{Thermally Comptonized Continuum}
\begin{footnotesize}
\begin{center}
\begin{tabular}{lll}
Parameter           &  Obs 1 (hard)                    &  Obs 8 (soft) \\
\tableline
\tableline
nH ($10^{22}$)      & 0.4*                        & 0.4* \\
kT (keV)            & 20*                      & $2.30^{+0.05}_{-0.02}$ \\
$\rm T_{0}$ (keV)   & $< 0.78$    & $0.49 \pm 0.2$ \\
$\rm \tau_{p}$      & $1.7\pm 0.1$        & $5.5 \pm 0.1$ \\
\tableline
approx   & 1     & 1 \\
\tableline
Flux $(10^{-9})$   & $3.5^{+0.6}_{-0.2}$ & \textbf{$4.78^{+0.05}_{-0.04}$}  \\
$\rm L/L_{Edd} (\%)$    & $7.7^{+1.3}_{-0.5}$        & $10.5^{+0.05}_{-0.05}$ \\
\tableline
$\chi^{2}/\nu$      &  115/107 &   419/424 \\
\tableline
\end{tabular}
\vspace*{\baselineskip}~\\
\end{center} 
\tablecomments{The results of fits to the hard and soft state spectrum of the neutron star of \gs\ with a Comptonization model. Parameters with * are frozen at the indicated values. The quoted errors are $1\sigma$ uncertainties. The continuum flux is quoted in units of ${\rm erg}~{\rm cm}^{-2}~{\rm s}^{-1}$ in the 0.1--100.0 keV~ band. }
\end{footnotesize}
\label{table:spectrum}
\end{table}

\subsection{Bolometric Flux and Eddington Luminosity}

In both cases we extrapolate the model over a 0.1 to 100 keV bandpass to estimate the bolometric flux. We stress that this results in a large degree of systematic uncertainty, especially in the hard state. In that case as we increase the (fixed) plasma kT, the overall bolometric flux also increases (because we are effectively adding flux outside of the bandpass). The flux level indicated in Table \ref{table:spectrum} should be considered as a lower limit, even though it is formally statistically constrained.

 We calculated the Eddington fraction using an inferred distance of $\left(5.7 \pm 0.2\right) \xi_b^{-1/2}$ kpc from \cite{Chenevez_2016}, where $\xi_b$ represents the possible anisotropy of the burst emission, and the Eddington luminosity of $\left(3.79 \pm 0.15\right) \times 10^{38} \ \rm erg \ s^{-1}$ calculated by \cite{Kuulkers_2003} for LMXBs with independently known distances. This gives an Eddington fraction of roughly 7\% for the hard state and roughly 10\% for the soft state, or a marginal increase in the inferred mass accretion rate.

\subsection{Thermonuclear Bursts}
\begin{figure*}[th]
\centering
\begin{minipage}{1.0\textwidth}
    \centering
    \includegraphics[width=1.0\textwidth]{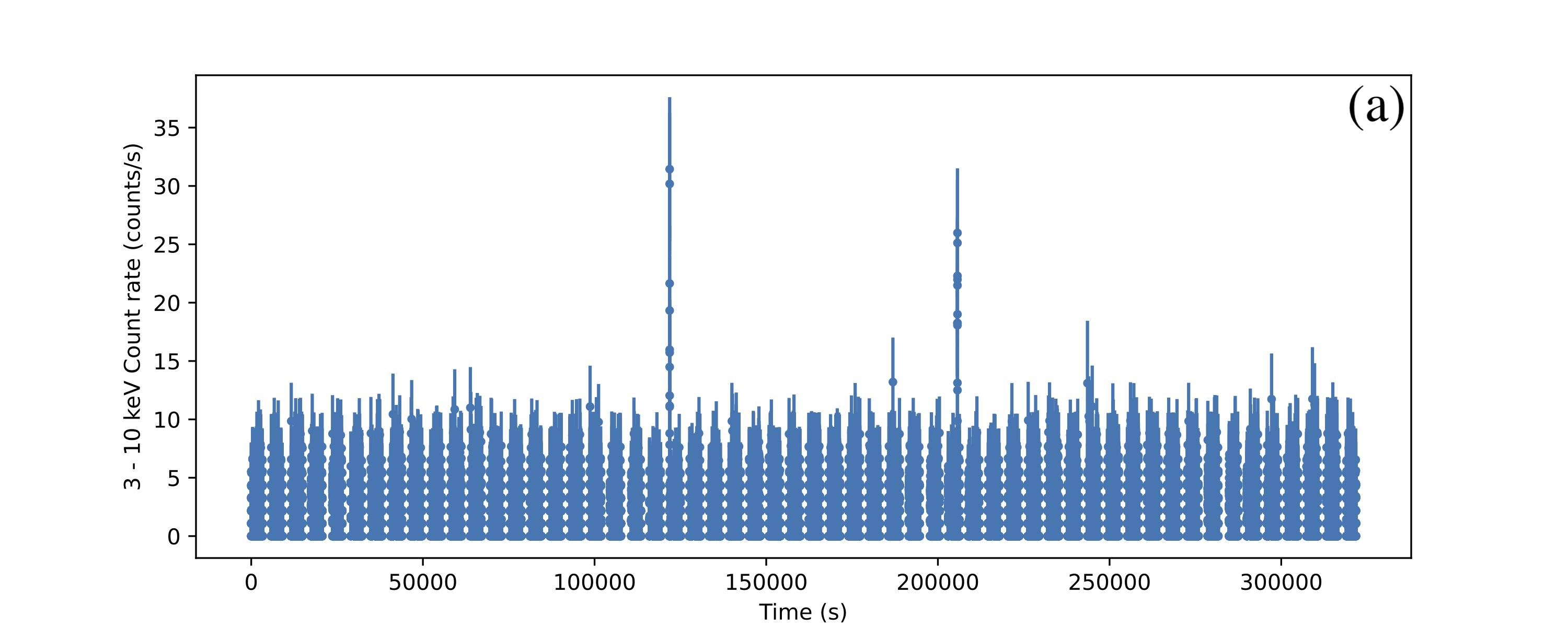}
\end{minipage}%
\hfill
\begin{minipage}{.47\textwidth}
    \centering
    \includegraphics[width=1.0\textwidth]{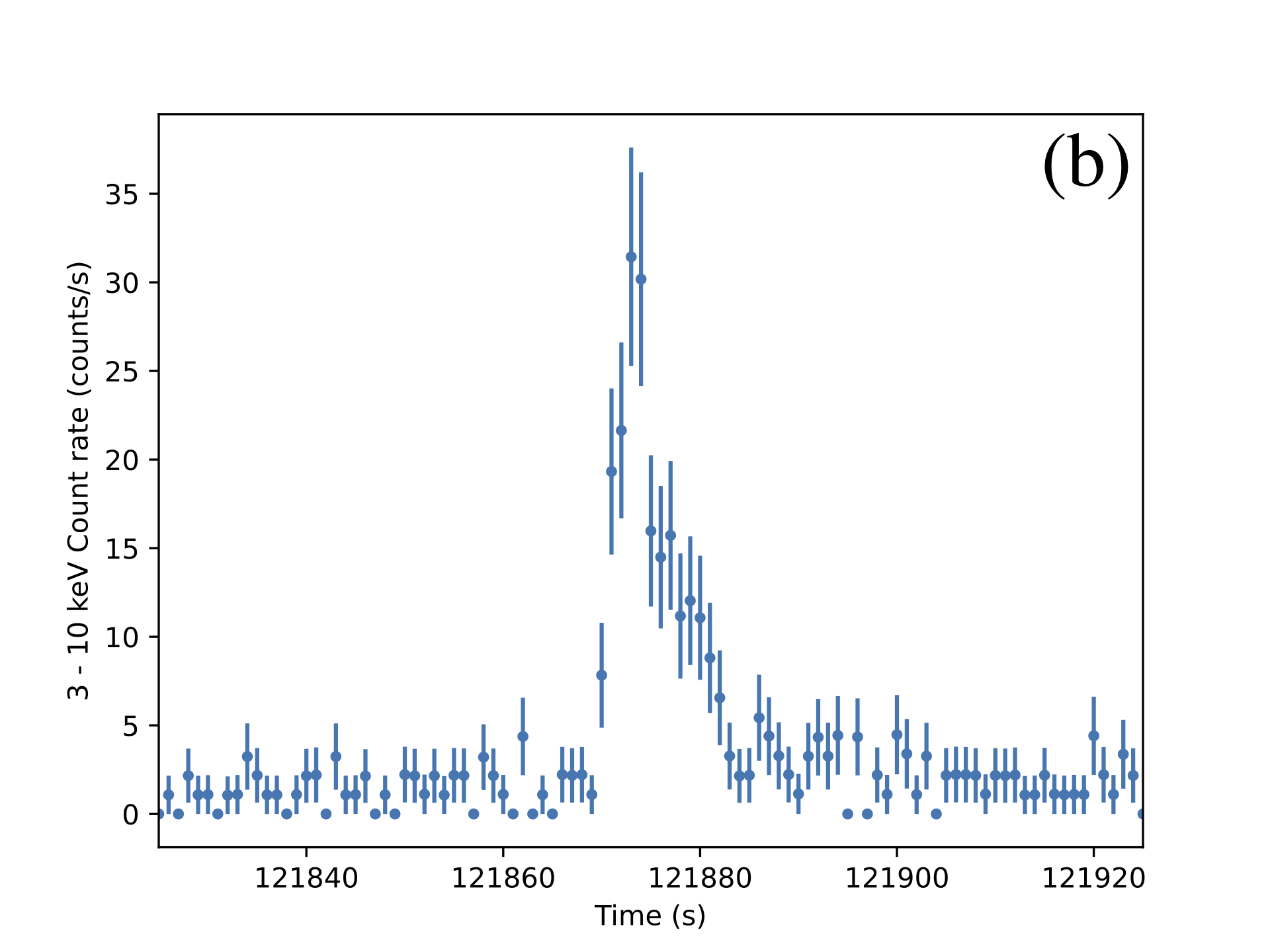}
\end{minipage}
\begin{minipage}{.47\textwidth}
    \centering
    \includegraphics[width=1.0\textwidth]{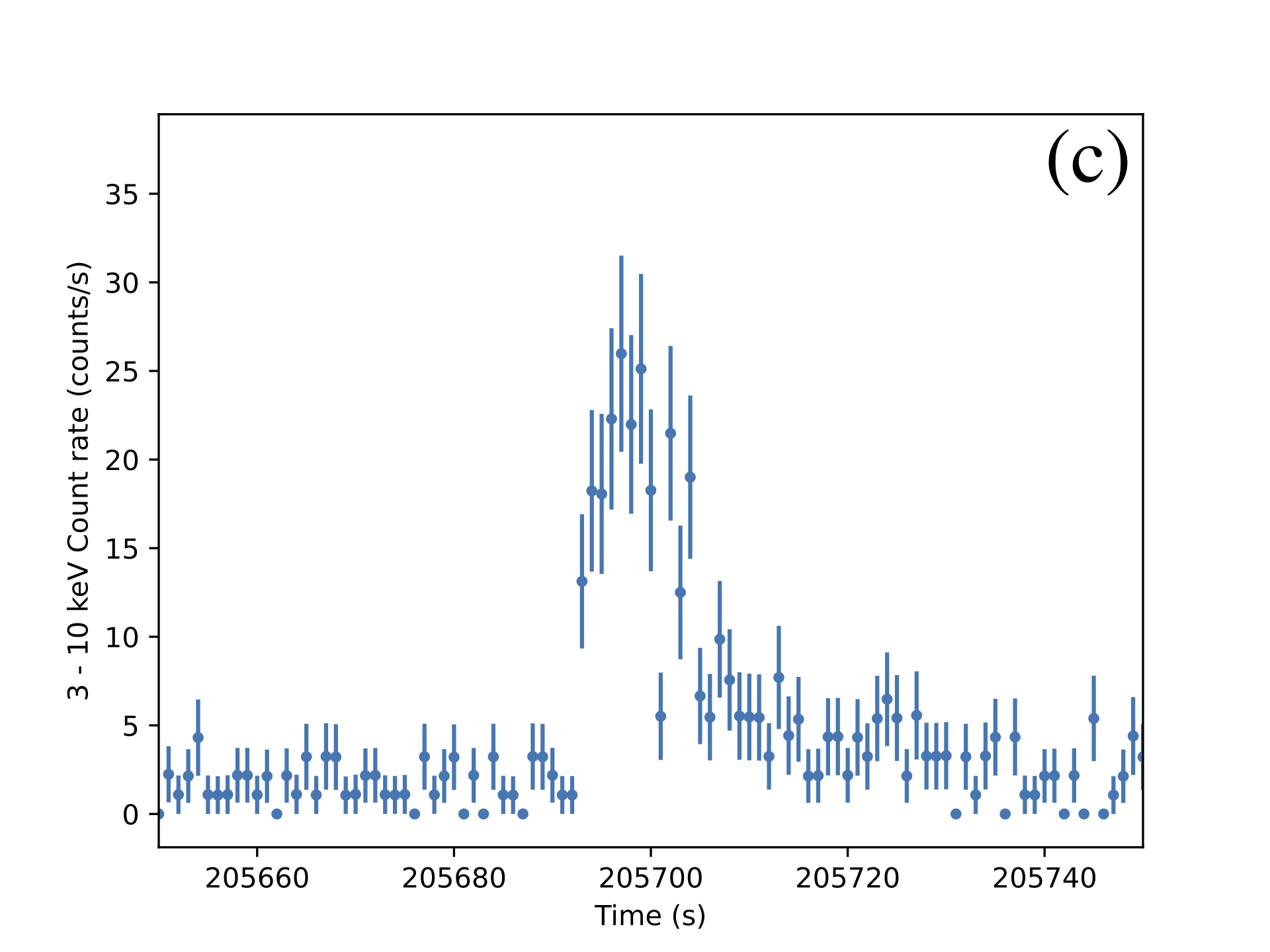}
\end{minipage}
\caption{(a) The total background subtracted, livetime corrected \nustar\ lightcurve of Obs 8 from 3-10 keV binned at time bins of 1s. The lightcurves (b) and (c) in the bottom panel are zoomed in to show the two Type 1 X-Ray bursts detected from this observation. }
\label{fig:total_lightcurve}
\end{figure*}

\begin{figure*}[t]
\centering
\begin{minipage}{.48\textwidth}
    \centering
    \includegraphics[width=1.0\textwidth]{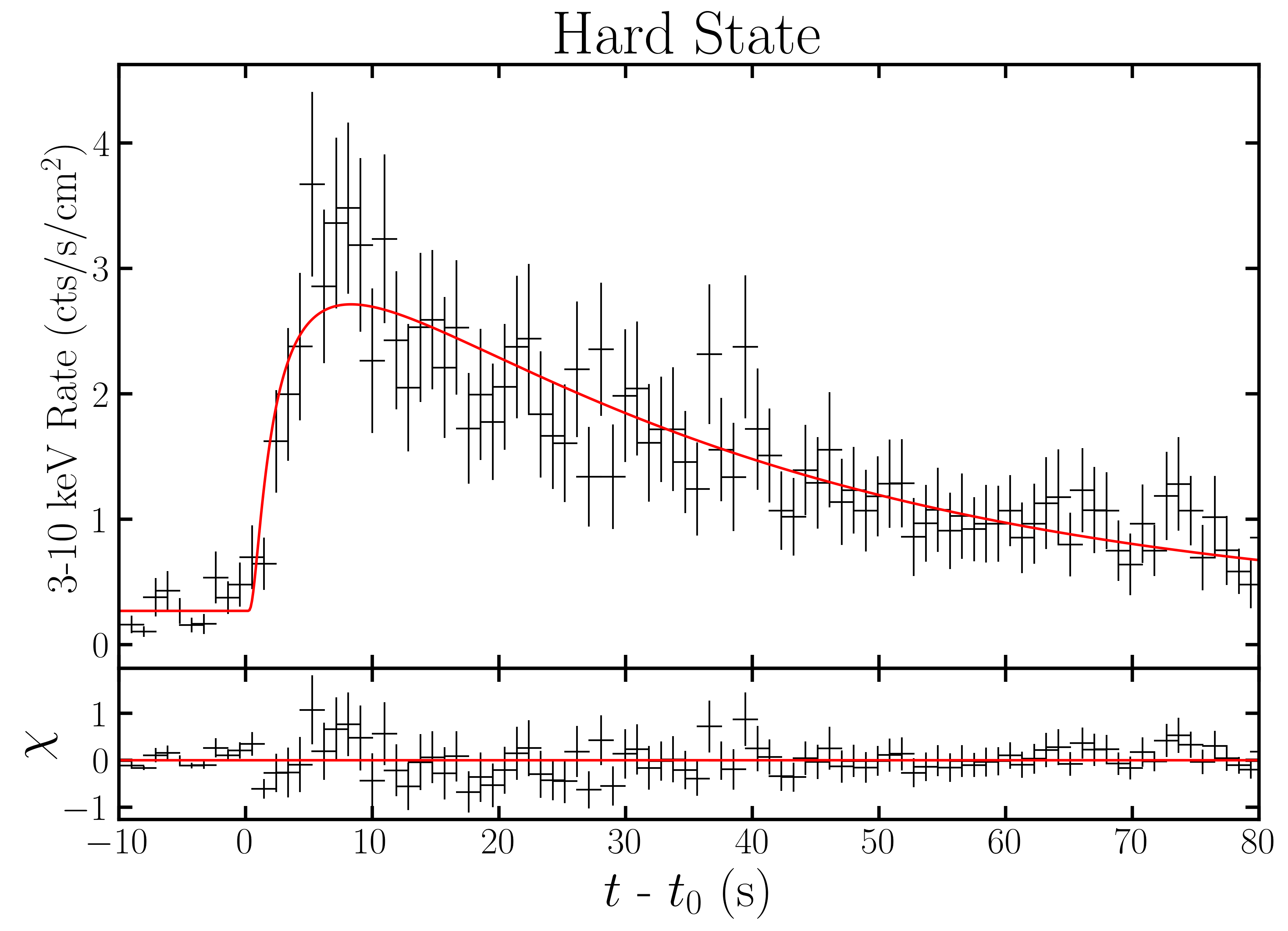}
\end{minipage}%
\hfill
\begin{minipage}{.48\textwidth}
    \centering
    \includegraphics[width=1.0\textwidth]{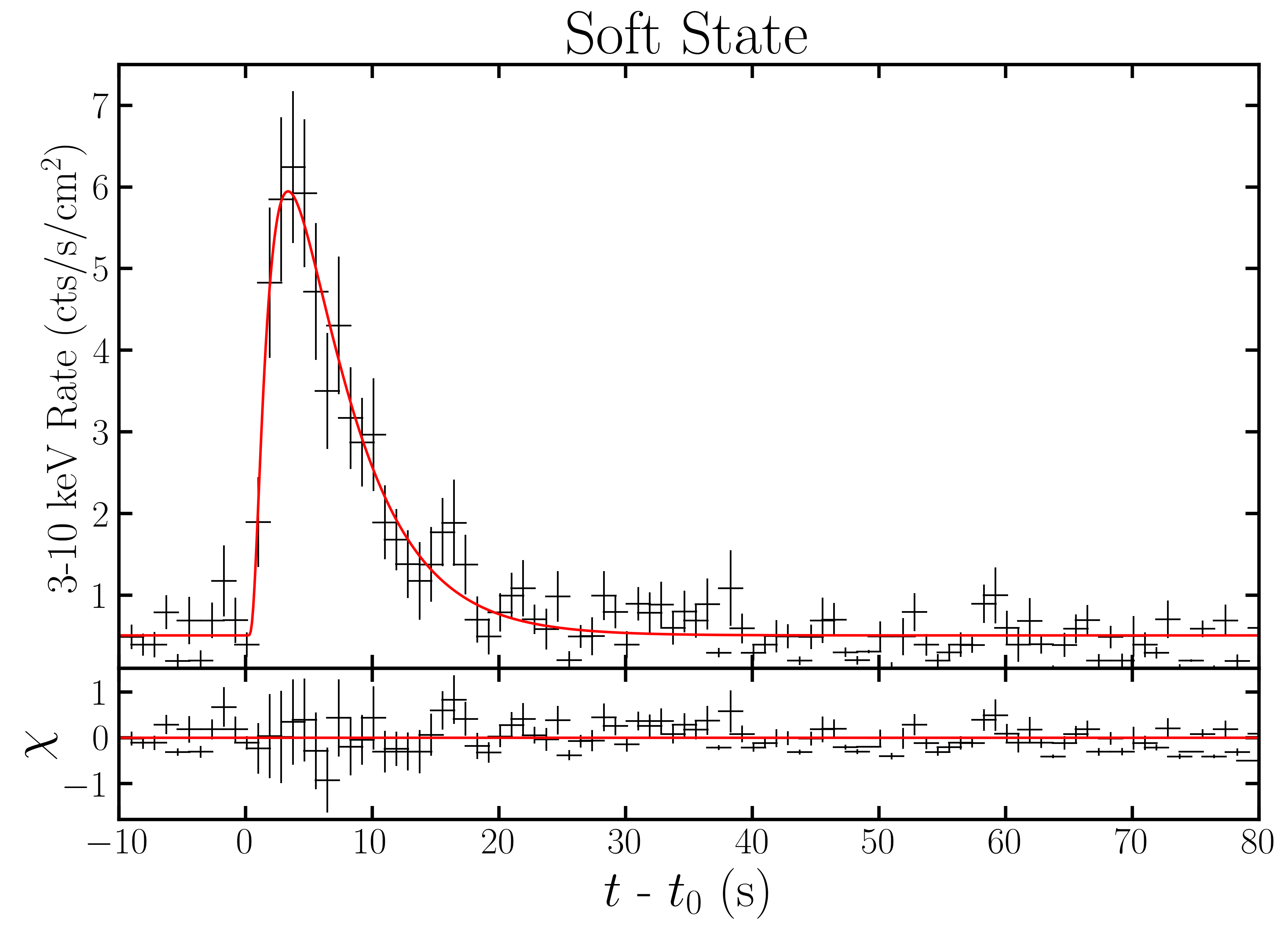}
\end{minipage}
\caption{The stacked burst light curves of the hard (\textit{Left}) and soft (\textit{Right}) epochs with their best-fit models and residuals (see text). The origin of the x-axis ($t - t_0 = 0 s$) corresponds to the time of burst onset, and the x-axis extends back to $t - t_0 = -10 s$ in order to illustrate the burst profile more clearly.}
\label{fig:lightcurve_model}
\end{figure*}

From the 3-10 keV lightcurves of the \nustar\ Stray Light observations of \gs, we saw little variability aside from occasional dramatic increases in count rates. As shown in Figure \ref{fig:total_lightcurve}, these features exhibit a fast rise followed by an exponential decay, which are characteristic behaviors of Type-1 X-ray bursts. We observed a total of 34 Type-I X-Ray bursts during our serendipitous observation of \gs \ using \nustar Stray Light. The number of bursts detected from each observation are listed Table \ref{table:Observation} and the zoomed in lightcurves of all the detected Type 1 bursts can be found in the \hyperref[sec:appendix]{Appendix}.

In order to better understand the structure of the Type-1 bursts and the differences of the bursts between spectral states, we fit the light curves to a simple Fast Rise Exponential Decay (FRED) model to better understand the structure of the two bursts. Due to the low count rates of our data, the Type-1 bursts have been stacked based on their respective spectral states. The start times for individual bursts were determined by fitting the model to each lightcurve. Then the light curves of each Type-1 burst were binned to 1s per time bin and stacked such that the start time ($t_0$) is set to zero. The FRED model was fit to the stacked burst lightcurves. The model is given by
\begin{equation}
     f(t) = A \: exp \left[ -\frac{\tau_R}{t-t_0} - \frac{t-t_0}{\tau_D} \right] + C
     \label{eqn:FRED}
\end{equation}
for $t > t_0$, where $t_0$ is the time of burst onset, $\tau_R$ and $\tau_D$ are the rise and decay times respectively, $A$ determines the height of the burst, and $C$ refers to the persistent count rate. From Equation \ref{eqn:FRED}, we can analytically compute the peak time of the burst:
\begin{equation}
    t_{peak} = \sqrt{\tau_R\tau_D} + t_0
    \label{eqn:t_peak}
\end{equation}
We also computed the time when the burst reaches the end of its tail, $t_{tail}$, defined as the time when the burst intensity drops to 25\% of its peak value. The total duration of the burst, $t_{90}$ was defined as the time between when the cumulative burst counts reaches 5\% and 95\% of the total integrated counts, allowing us to compute the average burst rate. The result of the fit and their residuals are shown in Figure \ref{fig:lightcurve_model} and Table \ref{table:lightcurve} shows the resultant fit parameters and calculations. 

\begin{table}[h]
\caption{Light Curve Fit Parameters}
\begin{footnotesize}
\begin{center}
\begin{tabular}{lll}
Parameter           &  Hard State                    &  Soft State \\
\tableline
\tableline
$\tau_R$ (s)              & $1.9 \pm 0.3$    & $2.5 \pm 0.2   $ \\
$\tau_D$ (s)              & $36 \pm 3$       & $4.6 \pm 0.3   $ \\
$A$ (cts $\rm s^{-1} \: cm^{-2}$)  & $3.9 \pm 0.2$    & $24 \pm 3    $ \\
$C$ (cts $\rm s^{-1} \: cm^{-2}$)  & $0.27 \pm 0.06$  & $0.51 \pm 0.04  $ \\
\tableline
$t_{peak}$ (s)            & $8.3 \pm 1.1$    & $3.3 \pm 0.3$ \\
$t_{tail} - t_{peak}$ (s) & $71.5 \pm 1.1$   & $10.4 \pm 0.3$ \\
$t_{90}$ (s)              & $105.2^{+7.8}_{-8.3}$   & $15.3^{+1.0}_{-0.9}$ \\
Integrated Counts (cts $\rm cm^{-2}$)& $133.8^{+2.5}_{-2.7}$   & $49.6^{+0.6}_{-0.4}$ \\
Avg. Burst Rate (cts $\rm s^{-1} \: cm^{-2}$)  & $1.27^{+0.13}_{-0.11}$  & $3.24^{+0.25}_{-0.23}$ \\
\tableline
\end{tabular}
\vspace*{\baselineskip}~\\
\end{center} 
\tablecomments{The results of fits to the lightcurves for the hard and soft state bursts. $t_{peak}$ refers to the time from the burst onset to the peak of the burst and $t_{tail}$ refers to the time when the intensity of the burst drops to 25\% of its peak value. The integrated counts are the counts integrated over the burst duration for each averaged lightcurve.}
\end{footnotesize}
\label{table:lightcurve}
\end{table}

The hard state burst shows close resemblance to a typical Type-1 burst, despite some residuals around the peak of the burst. The fit of the soft state bursts shows a clear residual. However it is not possible to determine if the fits indicate a double peaked structure or whether it is simply an artifact the stacking procedure. Comparing the bursts from the two spectral states, it is evident that the hard state bursts have a longer burst duration and has a larger burst fluence while the soft state bursts have a higher peak intensity and a higher average burst rate. The persistent count rate between bursts for the soft state is larger by a factor of two, consistent with our spectral fit results.

\section{Discussion}

\label{sec:discussion}

We confirm that \gs is an atoll source and was previously in the ``island" atoll state. Using spectra after the state transition seen in \maxi, our \nustar spectroscopic analysis confirms that the source is now in the ``banana" branch for an atoll source. The fact that the StrayCats data span both before and after the first coronal collapse event in 2014 and after the transitional period provides a unique view of the source through its transition.

The system took years to fully transition to the stable island state. With an orbital period of only 2.25-hr \citep{Homer_1998, Mescheryako_2010}, this implies some long-term instabilities in the accretion disk that are modulating the mass accretion rate and, therefore, the emergent X-ray spectrum. It's not clear what such a mechanism is, or what could trigger such long-term changes in behavior.

In neither state do we see clear evidence for line emission near 6 keV. This is not entirely unexpected given the SNR in Fig \ref{fig:spectral_fits}, we are not sensitive to weak Fe lines. A focused \nustar observation was obtained in \nustar AO Cycle 08 specifically to search for this feature (PI: Grefenstette) and will be reported on in future work.







\subsection{Change in the Type I X-ray Bursts}

Similar to the soft event from \cite{Chenevez_2016}, we see differences in the behavior of the Type I X-ray bursts between the two states (Table \ref{table:lightcurve}). In the hard state we see longer bursts with lower peak flux but higher burst fluence. The soft state has less regular, faster, and brighter burst with lower total integrated emission. Again, this may be the result of a change in the mass accretion rate and, possibly a chance in the composition of the accreted material \citep[see, e.g.,][]{Galloway2021}. Our interpretation is therefore that the increase in mass accretion rate associated with the transition to the brighter banana branch is consistent with the bursting behavior of the source.

In the hard state, the Type I bursts are indicative of He ignition in a Hydrogen rich environment, where nuclear burning via the rapid-proton process can lead to longer burst durations of $\approx100$s.
In the soft state, the short irregular bursts instead suggest that the accreted fuel is H-deficient and the resulting bursts largely caused by marginally stable He burning. This has been known to occur for mixed H/He accretors at high accretion rates, and is often accompanied by photospheric radius expansion and burst oscillations \citep{Galloway2021}. The physical mechanism behind the differences in burst behavior is yet unclear. Observations suggest that the disk geometry influences the burst mechanism and the accretion rate \citep{Galloway2021}. 


Compared to the hard state observations, where we mostly see at least 2-3 bursts per observation, there is a noticeable lack of bursts in the soft state. In particular, in only three observations (5A, 7, 8) do we see repeated bursts in the soft state.  This possibly indicates that the burst recurrence time has increased during the state transition. Obs 5 may be right on the transitional boundary. Obs 7 and 8 are the only soft state observations to contain multiple bursts and are likely only seen because of the long duration of the observations. It has been observed that bursts have relatively long or irregular recurrence times, with indications of stable burning in-between bursts, at the transition from a bursting to a stable burning regime \citep{Galloway2021}. Hence, this agrees with our previous statement about expecting a marginally stable He burning in the soft state.

Quasi-periodic oscillations (QPOs) at mHz frequencies are related to the occurrence of Type I bursts, and are thought to originate from oscillatory burning modes, resulting from marginally stable He burning \citep{Galloway2021}. \cite{Strohmayer_2018} reports observations of mHz oscillations made with \nicer, and indicates the possible presence of QPOs from the soft state of \gs. We were unable to search for such fast variability in this source due to the low SNR. However, we expect to be able to search for the mHz oscillations in the upcoming focused \nustar observation.

\section{Conclusions}

In this paper we have described \nustar StrayCats observations of \gs. We've discussed these observations in the context of the long-term behavior of the source as monitored by \swift-BAT and \maxi. The earliest observations find the source in a ``hard" state spectroscopically similar to other ``island" states in atoll sources; the late-time StrayCats observations find the source has transitioned to the ``banana" state. The Type I X-ray bursts likewise indicate the spectral state transition, and provides indication of the change in mass accretion rate and the fuel composition of the accretion disk of the source

\section*{Acknowledgements}

This work was supported by the National Aeronautics and Space Administration (NASA) under grant number 80NSSC19K1023 issued through the NNH18ZDA001N Astrophysics Data Analysis Program (ADAP). This research was performed at the California Institute of Technology (Caltech) under the Summer Undergraduate Research Fellowship (SURF) program.

Additionally, this work made use of data from the \nustar mission, a project led by the California Institute of Technology, managed by the Jet Propulsion Laboratory, and funded by the National Aeronautics and Space Administration. We thank the \nustar Operations, Software and Calibration teams for support with the execution and analysis of these observations. This research has made use of the \nustar Data Analysis Software (NuSTARDAS) jointly developed by the ASI Science Data Center (ASDC, Italy) and the California Institute of Technology (USA). This research has made use of data and/or software provided by the High Energy Astrophysics Science Archive Research Center (HEASARC), which is a service of the Astrophysics Science Division at NASA/GSFC.


\bibliography{references}
\bibliographystyle{aasjournal}

\newpage
\appendix \label{sec:appendix}

\begin{figure*}[h]\
\includegraphics[width=1.0\textwidth] {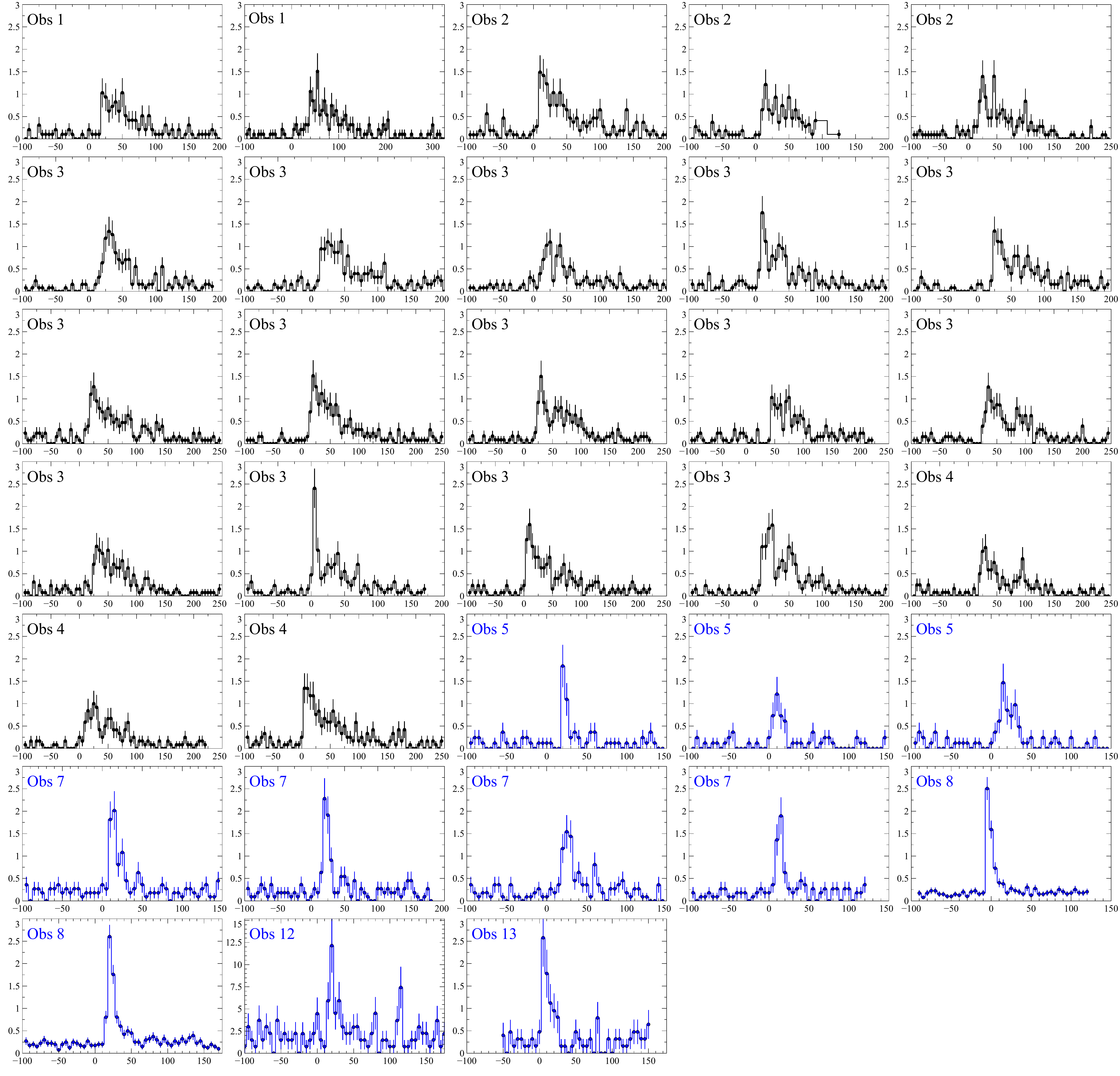}
\caption{A rogues gallery of the Type 1 X-Ray bursts observed in \gs. The bursts are color coded by their spectral states, black for the hard state bursts and blue for the soft state bursts. The rate per area is plotted against time, with \textit{t} = 0 being the approximate start times of the bursts. }
\label{fig:rogues}
\end{figure*}





\end{document}